\documentclass[pra,aps,twocolumn,10pt,superscriptaddress,notitlepage,longbibliography,groupedaddress]{revtex4-1}

\usepackage[margin=1.50cm]{geometry}
\usepackage[caption=false]{subfig}
\usepackage{graphicx}
\usepackage{amsmath,mathtools}
\usepackage{amssymb}
\usepackage{epstopdf}
\usepackage{siunitx}
\usepackage{color}
\usepackage[ngerman,english]{babel}
\usepackage{float}

\usepackage{layouts}
\graphicspath{{./Figs/}}

\usepackage{hyperref}
 \hypersetup{
     colorlinks=true,
     linkcolor=blue,
     filecolor=blue,
     citecolor = magenta,      
     urlcolor=red,
     }

\begin{document}
	
\title{Space-Time Computation and Visualization of the Electromagnetic Fields and Potentials Generated by Moving Point Charges}
\date{\today}
\author{Matthew J. Filipovich}
\email{matthew.filipovich@queensu.ca}
\affiliation{Department of Physics, Engineering Physics and Astronomy,
Queen's University, Kingston, ON K7L 3N6, Canada}
\author{Stephen Hughes}
\affiliation{Department of Physics, Engineering Physics and Astronomy,
Queen's University, Kingston, ON K7L 3N6, Canada}

\begin{abstract}
\noindent We present a computational methodology to directly calculate and visualize the directional components of the Coulomb, radiation, and total electromagnetic fields, as well as the scalar and vector potentials, generated by moving point charges in arbitrary motion with varying speeds. Our method explicitly calculates the retarded time of the point charge along a discretized grid which is then used to determine the fields and potentials. The computational approach, implemented in Python, provides an intuitive understanding of the electromagnetic waves generated by moving point charges and can be used as a pedagogical tool for undergraduate and graduate-level electromagnetic theory courses. Our computer code, freely available for download, can also approximate complicated time-varying continuous charge and current densities, and can be used in conjunction with grid-based numerical modeling methods to solve real-world computational electromagnetics problems, such as experiments with high-energy electron sources. We simulate and discuss several interesting example applications and lab experiments including electric and magnetic dipoles, oscillating and linear accelerating point charges, synchrotron radiation, and Bremsstrahlung. 
\end{abstract}

\maketitle

\section{Introduction}

The electric and magnetic fields generated by moving point charges are often complex and unexpected compared to their static counterparts.
In particular, for moving point charges, there is a ``correction'' to the retarded Coulomb field which is proportional to the rate of change of the retarded Coulomb field multiplied by the {\em retardation delay}, defined as the time it would take light to travel from the charge to the field point at speed $c$. If the charge is accelerating, there is an additional term in the electromagnetic (EM) field expressions, known as the radiation field, which is proportional to the second derivative of the vector directed from the charge at the retarded time towards the field point~\cite{feynman}. In homogeneous environments, the radiation field decreases as 1/$r$ from the point charge and is responsible for the EM radiation produced by moving charges that propagates to infinity. 

Retarded potentials and their solutions represent a special place in electrodynamics, generalizing
Poisson's equations to account for 
time dependence and retardation effects. Unlike Maxwell's equations,
the potentials have a freedom of gauge choice where the
potential need not satisfy causality, though the ``physical fields'' derived from these potential fields
must. Traditionally, such potentials are introduced in {\it advanced} or {\it retarded} forms, and involve
complicated integrations with respect to advanced or retarded time~\cite{Griffiths_2017}.
The potentials also appear naturally in quantum field theories, and some may argue they are more fundamental than the physical fields, explaining such peculiarities as the Aharonov–Bohm effect~\cite{PhysRev.115.485}. 
While elegant and insightful, there
are only a few known cases where an analytical solution
for such potentials exists.
These often involve fairly complex vector calculus and can leave many students struggling to visualize or appreciate the solution.
It is therefore unfortunate that few computational approaches have been developed to model such moving potential problems~\cite{Ruhlandt_Muhle_Enderlein_2020}.
Unlike many simulation tools with Maxwell's equations, such as the
finite-difference time-domain (FDTD) method,
there are no well known approaches to computationally solving problems with moving charges directly in space and time. 

In this paper, we introduce a direct numerical modelling technique to solve retarded time problems
of moving point charges. The simulation is implemented
in {\sc Python 3.8} for full three-dimensional problems and is optimized for computational efficiency. Our program allows the user to directly
visualize how the different EM fields and potentials are formed using numerical methods to determine the retarded time of the particle at different points in space.
Solutions of retarded potential
problems are not just an interesting academic study, but have direct relevance on emerging experiments with high-energy  electron sources, such
as electron-energy loss spectroscopy (EELS)~\cite{PhysRevLett.80.5180,PhysRev.100.1078}
and vortex EELS~\cite{Verbeeck2010,McMorran2011}.

The rest of our paper is organized as follows:
in Sec.~\ref{sec:theory}, we present the main theoretical background of moving point charges in electrodynamics, including the Li\'enard--Wiechert potentials~\cite{Wiechert_1901}
and expressions for the electric and magnetic fields.
In Sec.~\ref{sec:results}, we show a selection of results and examples that can be studied with our computational code, including the EM fields generated
from a fast oscillating charge and an oscillating dipole, as well as from steady state and time-varying current sources. We also explore the phenomena of synchrotron radiation, radiation emitted from a moving point charge with constant acceleration, and Bremsstrahlung.
We present our conclusions in
Sec.~\ref{sec:conclusions}.
The simulation code is
freely available for download and use at Ref.~\onlinecite{codes}.

\section{Theory and Computational Implementation}
\label{sec:theory}

The charge and current densities of a point charge $q$ at the position $\mathbf{r}_p(t)$ with velocity $c\boldsymbol{\beta}(t)$ are, respectively,
\begin{equation}
    \rho\left(\mathbf{r}, t\right) = q \delta\left[ \mathbf{r} - \mathbf{r}_p\left(t\right)\right],
\end{equation}
\begin{equation}
    \mathbf{J}\left(\mathbf{r}, t \right) = q c\boldsymbol{\beta}(t) \delta \left[ \mathbf{r} - \mathbf{r}_p\left(t\right)\right].
\end{equation}
The scalar and vector potentials of a moving point charge in the Lorenz gauge, known as the Liénard–Wiechert potentials~\cite{Wiechert_1901},  are derived from Maxwell's equations as
\begin{equation}\label{V}
    \Phi(\mathbf{r}, t) = \frac{q}{4\pi\epsilon_0}\left[ \frac{1}{\kappa R}\right]_{\mathrm{ret}},
\end{equation}
\begin{equation}\label{A}
    \mathbf{A}(\mathbf{r}, t) = \frac{\mu_0 q}{4\pi}\left[ \frac{\boldsymbol{\beta}}{\kappa R}\right]_{\mathrm{ret}},
\end{equation}
where $R=|\mathbf{r}-\mathbf{r}_p(t')|$, $\kappa=1-\mathbf{n}(t')\cdot \boldsymbol{\beta}(t')$ such that ${\mathbf{n}=(\mathbf{r}-\mathbf{r}_p(t'))/R}$ is a unit vector from the position of the charge to the field point, and the quantity in brackets is to be evaluated at the {\it retarded time} $t'$ given by
\begin{equation}\label{t}
    t' = t-\frac{R(t')}{c}.
\end{equation}

The physical (gauge-invariant) electric and magnetic fields generated from a moving point charge can be obtained using various approaches, including deriving them directly from their scalar and vector potentials~\cite{Griffiths_2017, Jackson_1999}:
\begin{equation}\label{E}
    \mathbf{E}\left(\mathbf{r}, t\right) = \frac{q}{4\pi\epsilon_0} \Bigg[ \frac{\left( \mathbf{n}-\boldsymbol{\beta} \right)\left(1-\beta^2\right)}{\kappa^3 R^2} + \frac{\mathbf{n}}{c\kappa^3 R} \times \left[ \left(\mathbf{n}-\boldsymbol{\beta}\right) \times \boldsymbol{\dot{\beta}} \right] \Bigg]_{\mathrm{ret}},
\end{equation}
\begin{equation}\label{B}
    \mathbf{B}\left(\mathbf{r}, t\right) = \frac{1}{c} \left[ \mathbf{n} \times \mathbf{E} \right]_{\mathrm{ret}},
\end{equation}
where $\boldsymbol{\dot{\beta}}$ is the derivative of $\boldsymbol{\beta}$ with respect to $t'$. The first term in Eq.~\eqref{E} is known as the electric Coulomb field and is independent of acceleration, while the second term is known as the electric radiation field and is linearly dependent on $\boldsymbol{\dot{\beta}}$:
\begin{equation}\label{ECoulomb}
    \mathbf{E}_{\mathrm{Coul}}\left(\mathbf{r}, t\right) = \frac{q}{4\pi \epsilon_0}\left[\frac{\left(\mathbf{n}-\boldsymbol{\beta} \right)\left(1-\beta^2\right)}{\kappa^3 R^2}  \right]_{\mathrm{ret}},
\end{equation}
\begin{equation}\label{Eradiation}
    \mathbf{E}_{\mathrm{rad}}\left(\mathbf{r}, t\right) = \frac{q}{4\pi\epsilon_0c}\left[\frac{\mathbf{n}}{\kappa^3R}\times \left[ \left( \mathbf{n} - \boldsymbol{\beta} \right) \times \boldsymbol{\dot{\beta}}\right] \right]_{\mathrm{ret}}.
\end{equation}
The magnetic Coulomb and radiation field terms can be determined by substituting Eqs.~\eqref{ECoulomb} and~\eqref{Eradiation} into Eq.~\eqref{B}. Notably, the Coulomb field falls off as $1/R^{2}$, similar to the static field, while the radiation field decreases as $1/R$.

To computationally solve the above equations, 
our simulation determines the retarded time $t'$ of a moving point charge in arbitrary motion at each time step using Newton's method to calculate the approximate solution of Eq.~\eqref{t} for~$t'$. For the simulations, the trajectories of the moving point charges, including the velocities and accelerations at each time step, are known {\it a priori}. We simulate an arbitrary number of point charges by exploiting the superposition principle for EM  fields and potentials. In addition to specifying the particles' trajectories beforehand, their motion could be determined numerically at each time step using the Lorentz force: $\mathbf{F}=q\mathbf{E}+q\mathbf{v}\times\mathbf{B}$, where $\mathbf{v}$ is the instantaneous velocity of the point charge and the $\mathbf{E}$ and $\mathbf{B}$ fields are generated from other charges in the simulation.

The scalar and vector potentials are calculated at each time step from Eqs.~\eqref{V}~and~\eqref{A} using the previously determined retarded times at each grid point. The total, Coulomb, and radiation fields are computed using Eqs.~\eqref{E},~\eqref{ECoulomb}, and~\eqref{Eradiation} for the respective electric fields; the corresponding magnetic fields are calculated from Eq.~\eqref{B}. 

A continuous charge density $\rho$ can be approximated in the simulation using numerous point charges within the volume, where the charge value of each point charge depends on $\rho$. Similarly, a continuous current density, described by $\mathbf{J}=\rho \mathbf{v}$, can be approximated in the simulation using evenly spaced point charges traveling along a path where the charge value of each point charge depends on $\mathbf{J}$. The accuracy of the calculated fields and potentials generated by these approximated continuous densities is dependent on both the number of point charges used in the simulation and the distance at the field point from the point charges.

Our open-source electromagnetic simulator was written in {\sc Python~3.8} and is available at Ref.~\onlinecite{codes}, which includes the code used to produce the figures in this paper and the corresponding figure animations, as well as a tutorial written in Jupyter Notebook that demonstrates the creation and visualization of custom simulations. The program allows the user to specify the charge and trajectory of each point charge in the simulation. The code is optimized using vectorized operations from the 
{\sc NumPy} package to compute the retarded time of the particles, as well as the potentials and fields, along the grid. The program runs $\mathcal{O}(n)$ with respect to both the number of particles in the simulation and the number of grid points. Using an Intel Core i5-8250U 1.60~GHz CPU, the electric and magnetic fields generated by a single moving point charge can be calculated at each point in a three dimensional grid of size $100\times100\times100$ in approximately 
2.1 seconds.

The figures shown in Sec.~\ref{sec:results} are the steady state solutions of the EM  fields generated from various moving point charge configurations. The point charges in the plots below carry a charge $e$ unless stated otherwise. These snapshots were captured after a sufficient  time duration had elapsed since the point charges began moving from their initial stationary state. The transient behaviour of the fields can be seen in the accompanying animations. 

\section{Results}
\label{sec:results}

\subsection{Oscillating Single Charge}
\label{sec:oscil_single}

\begin{figure}
    \centering
    \includegraphics{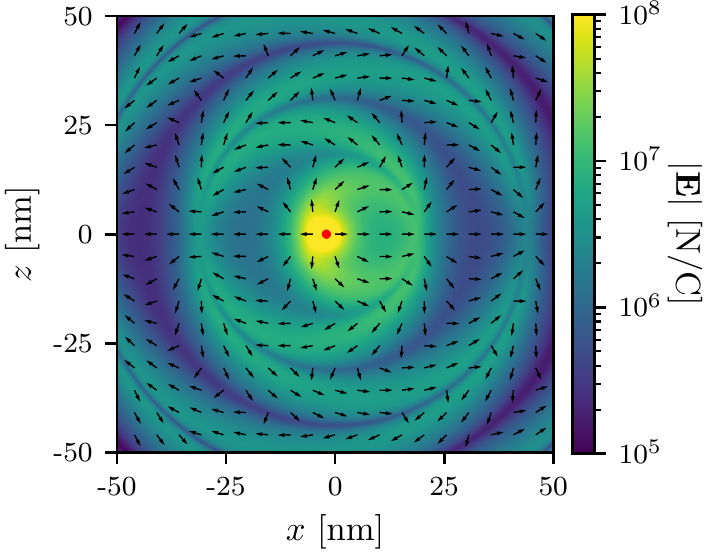}
    \caption{Magnitude of the total electric field in the $xz$ plane from a sinusoidal oscillating positive point charge with arrows representing the normalized direction of the field. The point charge is oscillating along the $x$ axis with an amplitude of 2~nm and a maximum point charge speed of $0.5c$, yielding an angular frequency $\omega\simeq 7.495\times 10^{16}$~rad/s. The position of the charge is shown by the red dot at $x=-2$ nm.}
    \label{fig:xz_sin_direction}
\end{figure}

\begin{figure}
    \centering
    \includegraphics[width=\linewidth]{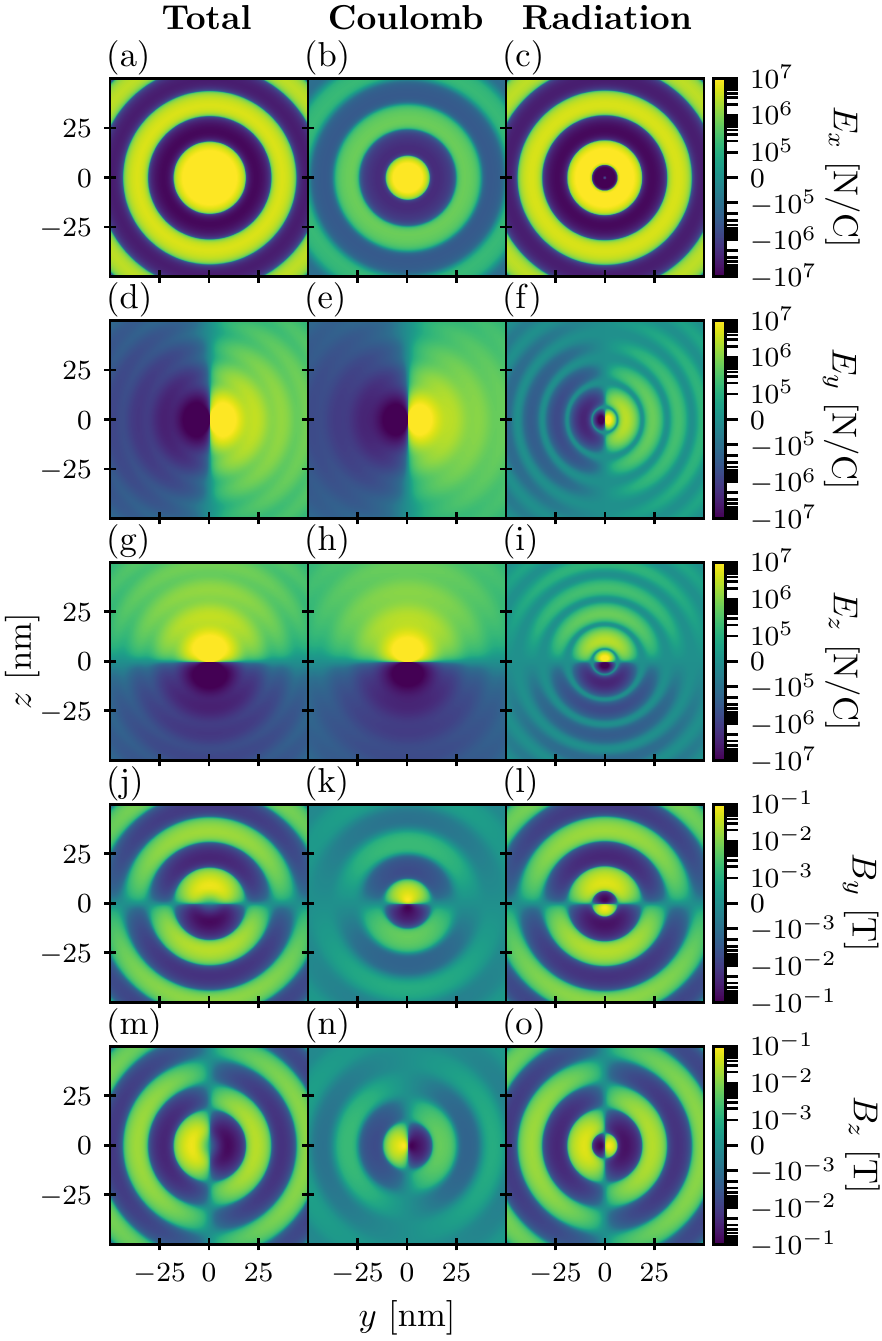}
    \caption{EM  fields from a sinusoidal oscillating positive point charge along the $x$ axis with an amplitude of 2~nm and a maximum point charge speed of $0.5c$. Snapshots in the $yz$ plane where the charge is at position ${x=-2~\mathrm{nm}}$. The total, Coulomb, and radiation fields are plotted from left to right: (a)--(c)~$E_x$; (d)--(f)~$E_y$; (g)--(i)~$E_z$; (j)--(l)~$B_y$; (m)--(o)~$B_z$.}
    \label{fig:yz_sin_field}
\end{figure}

\begin{figure}
    \centering
    \includegraphics[width=\linewidth]{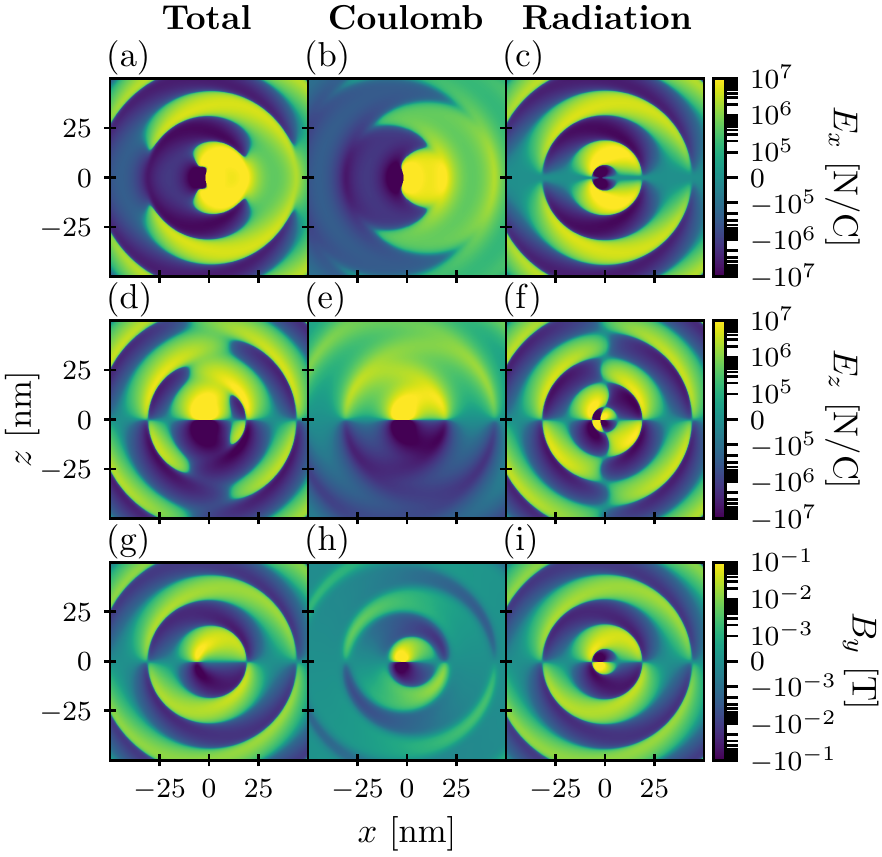}
    \caption{EM  fields in the $xz$ plane from a sinusoidal oscillating positive point charge as shown in Fig.~\ref{fig:yz_sin_field}. The total, Coulomb, and radiation fields are plotted from left to right: (a)--(c)~$E_x$; (d)--(f)~$E_z$; (g)--(i)~$B_y$.}
    \label{fig:xz_sin_field}
 \end{figure}

\begin{figure}
    \centering
    \includegraphics{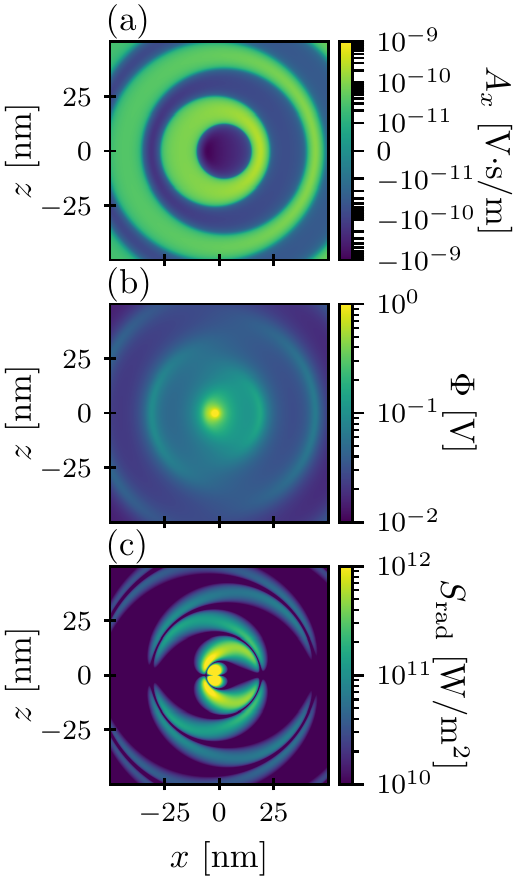}
    \caption{Potentials and Poynting vector in the $xz$ plane of a sinusoidal oscillating positive point charge as shown in Fig.~\ref{fig:yz_sin_field}. (a)~$x$ component of the vector potential; (b)~scalar potential; (c)~Poynting vector for the radiation field in radial direction $\mathbf{\hat{r}}$.
    }
    \label{fig:xz_sin_pot}
\end{figure}

First, we simulate a single positive charge oscillating sinusoidally along the $x$ axis that reaches a maximum speed of $0.5c$ and has an amplitude $A=2$ nm. The trajectory of the point charge is described by ${x(t)=A \cos(\omega t)}$, where ${\omega=0.5c/A}$. At this speed, the particle produces radiation with a wavelength $\lambda\simeq25.133$~nm. To illustrate the vectorial properties of the fields, the magnitude of the total electric field with arrows representing the field's direction in the $xz$ plane is shown in Fig.~\ref{fig:xz_sin_direction}. The components of the total (left), Coulomb (center), and radiation (right) fields produced by the accelerating particle are shown in the $yz$ and $xz$ planes in Figs.~\ref{fig:yz_sin_field} and~\ref{fig:xz_sin_field}, respectively. The total field comprises both the Coulomb and radiation field, and only the EM field components with non-zero values are shown. The plots use a symmetric logarithmic scale that is linear around zero to allow both positive and negative values. The scalar and vector potentials, as well as the Poynting vector for the radiation field $\mathbf{S}_{\mathrm{rad}}=\frac{1}{\mu_0}\left(\mathbf{E}_{\mathrm{rad}}\times\mathbf{B}_{\mathrm{rad}}\right)$, are shown in the $xz$ plane in Fig.~\ref{fig:xz_sin_pot}.

The Coulomb and radiation fields generated by the oscillating point charge are comparable in magnitude at the length scale shown in Figs.~\ref{fig:yz_sin_field} and \ref{fig:xz_sin_field}, and therefore the contributions from both are significant for the total field. However, at larger distances from the point charge, the contribution from the Coulomb field becomes negligible due to its inverse-square relation with distance, as the radiation field is only inversely proportional. In addition, at non-relativistic speeds, the contribution from the radiation field is negligible at this length scale.

The components of the electric Coulomb field shown in the plots resemble the static point charge field, described by Coulomb's law, propagating outwards at the speed of light. The wave nature of the fields is apparent in the corresponding animation. However, the shape of the electric radiation field is unexpected compared to its static counterpart. Unlike the electric Coulomb field, the electric radiation field in the direction of acceleration is zero, as shown in Fig.~\ref{fig:xz_sin_field}~(c) and~(f). Therefore, there is no power radiated in the direction of acceleration, as shown in Fig.~\ref{fig:xz_sin_pot}~(c). The planes of symmetry for the electric Coulomb and radiation fields are also different.

\subsection{Oscillating Electric Dipole}

Electric dipole radiation is usually responsible for the radiation emitted in the electronic transitions of atoms or molecules, and can describe the scattered light by a nanoparticle in a laser beam to the first approximation~\cite{Li_Arnoldus_2010}. The solutions for problems involving oscillating physical (finite-size) electric dipoles are typically approximated using multipole expansions, as well as by separating the regions of the EM field into the ``near field'' and ``far field.''
However, for small particles (much smaller than the excitation wavelength), the dipole approximation is known to be excellent. Using our simulation, we are able to calculate the exact numerical solution for the potentials and fields at all points in space.

\begin{figure}
    \centering
    \includegraphics[width=\linewidth]{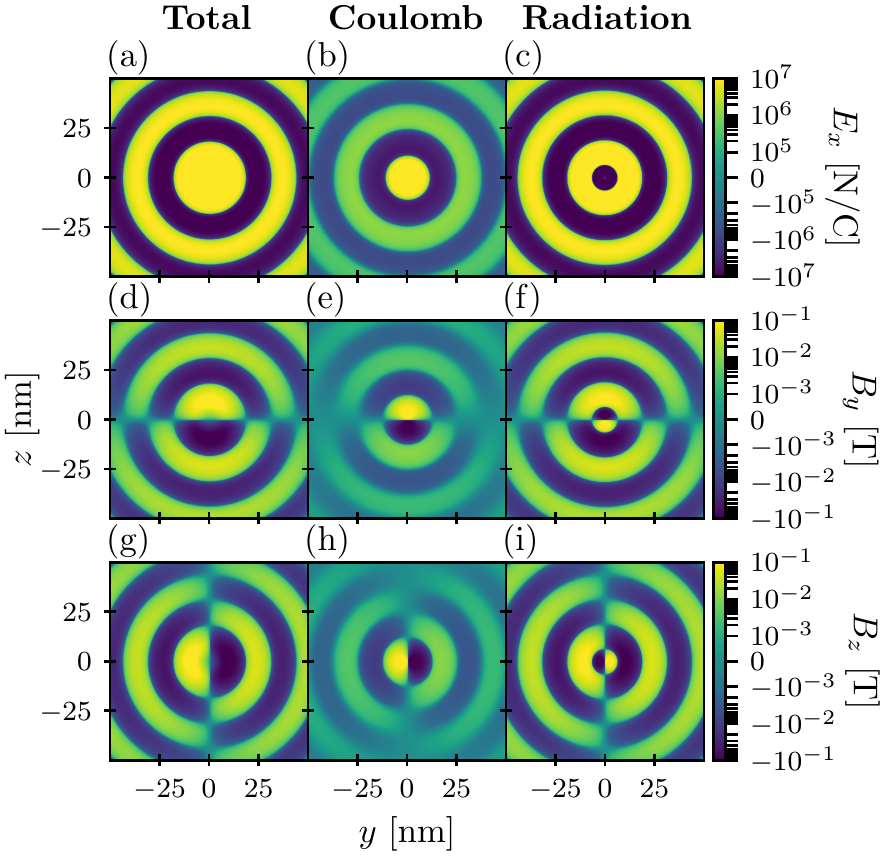}
    \caption{EM fields from a sinusoidal oscillating electric dipole along the $x$ axis with an amplitude of 2 nm and a maximum point charge speed of $0.5c$, yielding an angular frequency $\omega\simeq7.495\times 10^{16}~\mathrm{rad/s}$. Snapshots in the $yz$ plane where the positive charge is at position ${x=-2~\mathrm{nm}}$ and the negative charge is at ${x=2~\mathrm{nm}}$. The total, Coulomb, and radiation fields are plotted from left to right: (a)--(c) $E_x$; (d)--(f) $B_y$; (g)--(i)~$B_z$.
    }
    \label{fig:yz_dipole_field}
\end{figure}

\begin{figure}
    \centering
    \includegraphics[width=\linewidth]{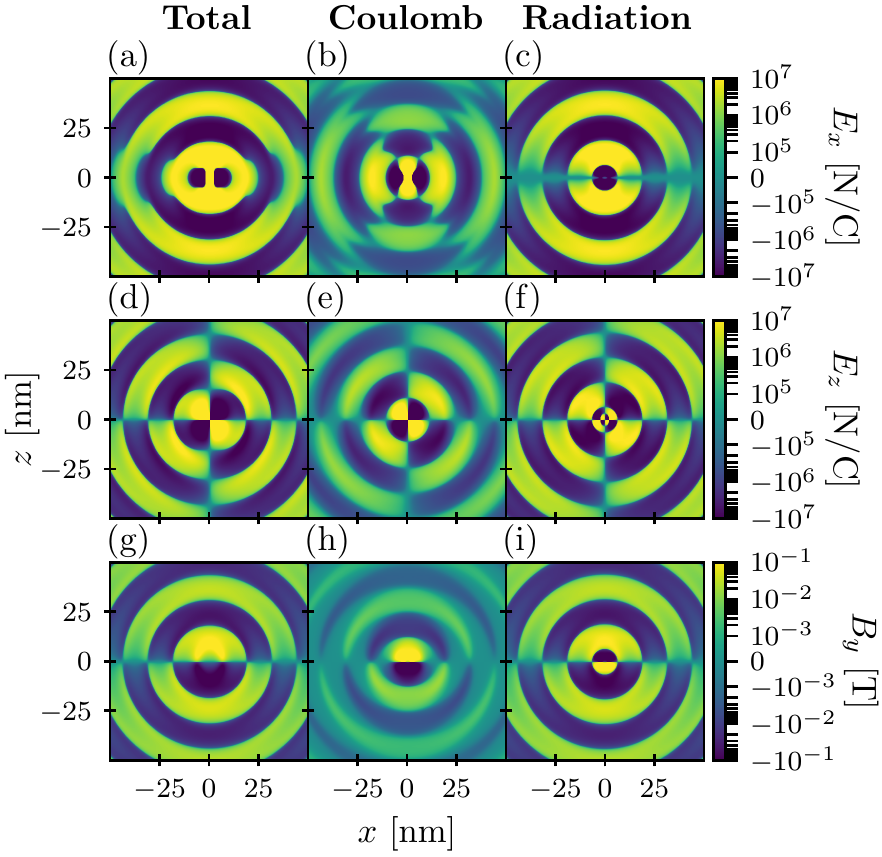}
    \caption{EM fields in the $xz$ plane from a sinusoidal oscillating electric dipole as shown in Fig.~\ref{fig:yz_dipole_field}. The total, Coulomb, and radiation fields are plotted from left to right: (a)--(c)~$E_x$; (d)--(f)~$E_z$; (g)--(i)~$B_y$.}
    \label{fig:xz_dipole_field}
    \vspace{2pt}
    \centering
    \includegraphics{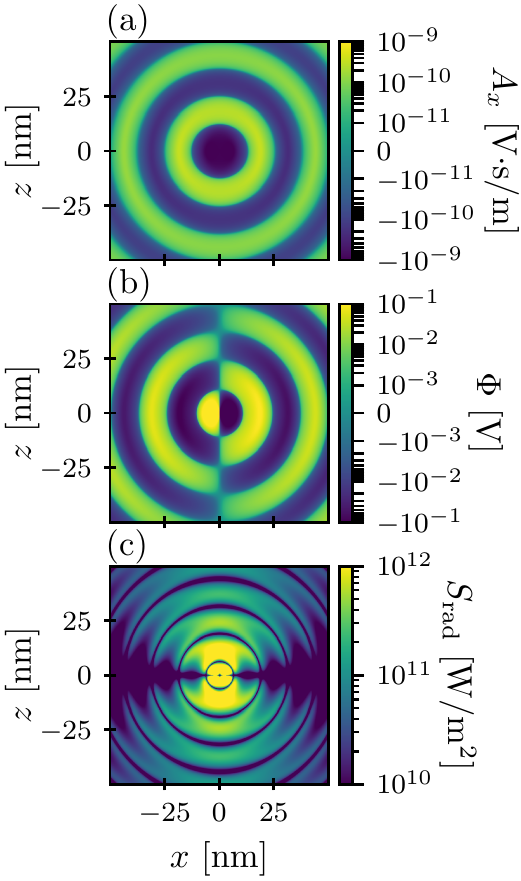}
    \caption{Potentials and Poynting vector in the $xz$ plane of a sinusoidal oscillating electric dipole as shown in Fig.~\ref{fig:yz_dipole_field}. (a)~$x$ component of the vector potential; (b)~scalar potential; (c)~Poynting vector for the radiation field in radial direction $\mathbf{\hat{r}}$.}
    \label{fig:xz_dipole_pot}
\end{figure}

Here, we simulate a physical dipole oscillating sinusoidally along the $x$ axis with a maximum charge separation of 4 nm and a maximum particle speed of $0.5c$, similar to the single oscillating charge show in Sec.~\ref{sec:oscil_single}. The EM  fields generated from the charges are shown in the $yz$ and $xz$ plane in Figs.~\ref{fig:yz_dipole_field} and~\ref{fig:xz_dipole_field}, respectively. The potentials and Poynting vector for the radiation field are shown in the $xz$ plane in Fig.~\ref{fig:xz_dipole_pot}. As expected from the analytical solutions for a moving dipole, there is no radiation emitted along the axis of the dipole.

The general shape of the electric radiation field emitted from the moving dipole is similar to the single oscillating charge, while the electric Coulomb fields along the $xz$ axis are distinct. The differences in the fields are caused by the constructive and destructive interference of the EM waves generated by the positive and negative oscillating charges. The results for an oscillating electric dipole can be extrapolated to describe the radiation emitted from a dipole antenna, and our simulation can be used to calculate the exact theoretical radiation pattern from phased array antennas.

\subsection{Synchrotron Radiation}

Synchrotron radiation is produced when charged particles are accelerated radially. This can be achieved in synchrotrons, which are high-energy particle machines that inject charged particles into roughly circular orbits. Synchrotrons behave like huge excited antennas radiating EM energy with a  broad spectrum that contains the frequency of revolution and the corresponding harmonics~\cite{kunz1974synchrotron}.

\begin{figure}
    \centering
   \includegraphics[width=\linewidth]{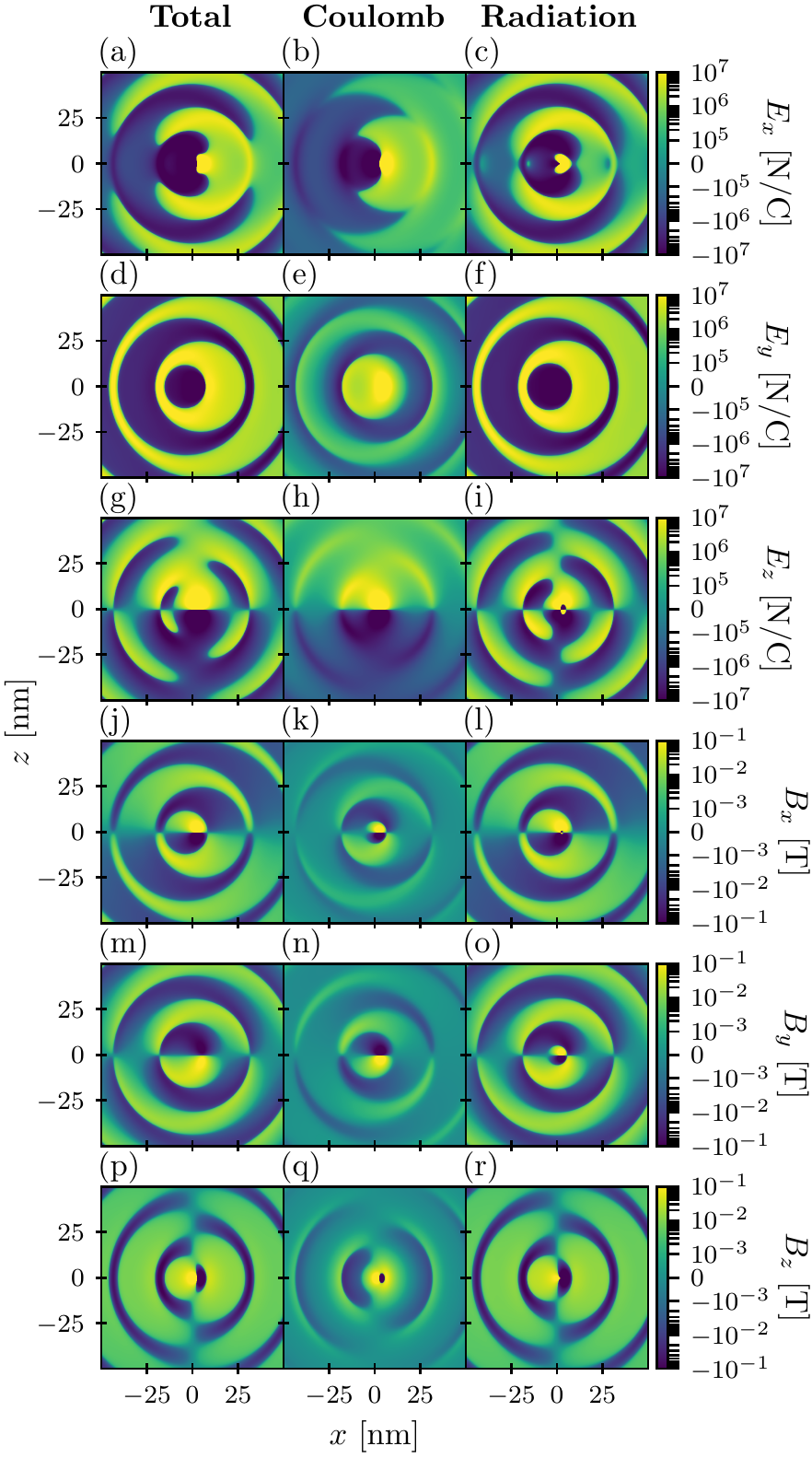}
    \caption{EM fields from a radially accelerating positive point charge orbiting in the $xy$ plane with a radius of 2 nm and a speed of $0.5c$, yielding an angular frequency $\omega\simeq7.495\times 10^{16}$~rad/s. Snapshots in the $xz$ plane where the charge is at position ${x=2~\mathrm{nm}}$ and ${y=0~\mathrm{nm}}$. The total, Coulomb, and radiation fields are plotted from left to right: (a)--(c)~$E_x$; (d)--(f)~$E_y$; (g)--(i)~$E_z$; (j)--(l)~$B_x$; (m)--(o)~$B_y$; (p)--(r)~$B_z$.}
    \label{fig:xz_radial_field}
\end{figure}

\begin{figure}
    \centering
    \includegraphics[width=\linewidth]{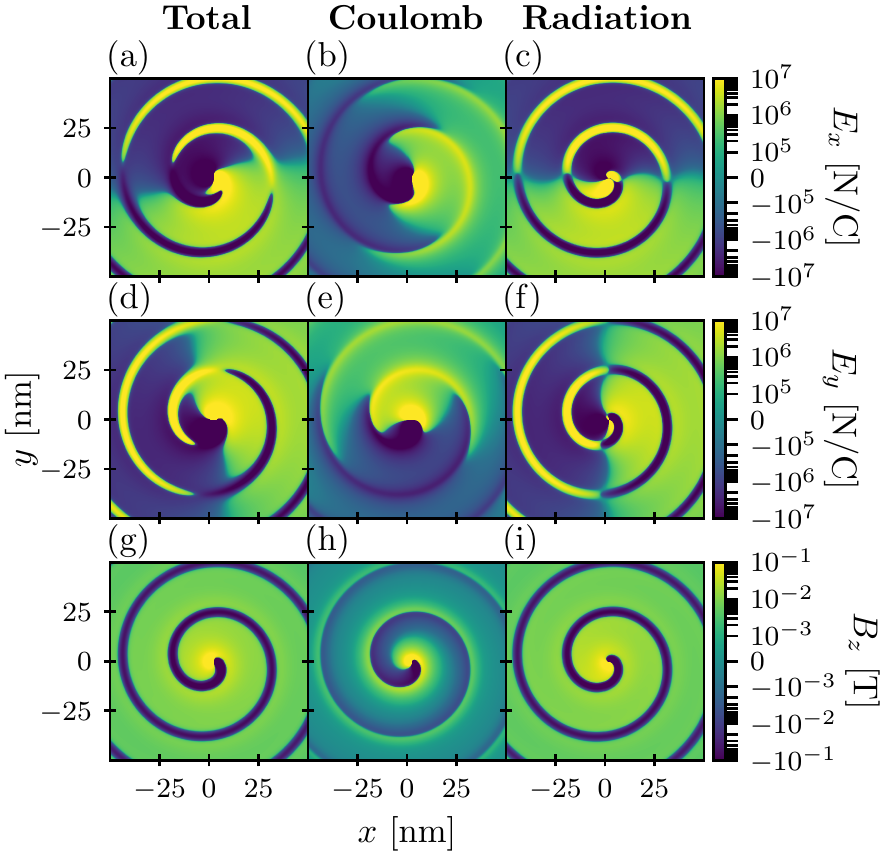}
    \caption{EM fields in the $xy$ plane from a radially accelerating positive point charge as shown in Fig.~\ref{fig:xz_radial_field}. The total, Coulomb, and radiation fields are plotted from left to right: (a)--(c)~$E_x$; (d)--(f)~$E_y$; (g)--(i)~$B_z$.}
    \label{fig:xy_radial_field}
\end{figure}

\begin{figure}
\includegraphics[width=\linewidth]{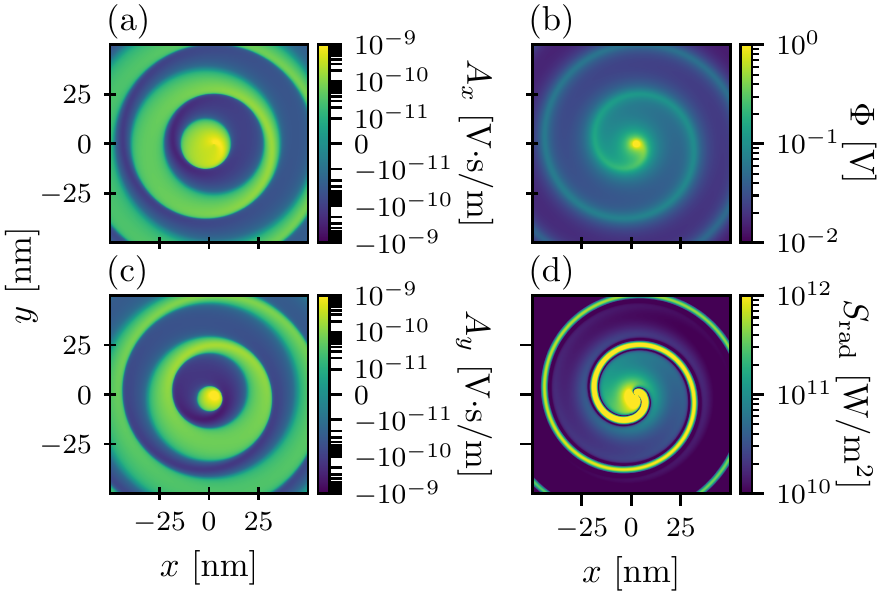}
    \centering
    \caption{Potentials and Poynting vector in the $xy$ plane of a radially accelerating positive point charge as shown in Fig.~\ref{fig:xz_radial_field}. (a)~$x$ component of the vector potential; (b)~$y$ component of the vector potential; (c)~scalar potential; (d)~Poynting vector for the radiation field in radial direction~$\mathbf{\hat{r}}$.}
    \label{fig:xy_radial_pot}
\end{figure}

We directly simulate a radially accelerating positive point charge that follows a circular trajectory with a radius of 2~nm and a speed of $0.5c$ along the $xy$ plane. The EM fields in the $xz$ and $xy$ planes are shown in Figs.~\ref{fig:xz_radial_field} and~\ref{fig:xy_radial_field}, respectively, and the potentials and Poynting vector for the radiation field in the $xy$ plane are shown in Fig.~\ref{fig:xy_radial_pot}.
The $x$ and $z$ components of the electric field along the $xz$ plane are very similar in shape to the components of the field generated by the single oscillating charge shown in Fig.~\ref{fig:xz_sin_field}. The similarity exists between these fields because in both cases, the point charge oscillates perpendicular to the $xz$ plane. However, the EM fields generated from the radially accelerating point charge along the $xy$ plane follow a spiral pattern that propagates radially outwards for both the Coulomb and radiation fields. This spiral pattern is also shown in the Poynting vector plot, where the peak energy flux is along the spiral radiating from the point charge.

\subsection{Current Loops}

Point charges can be used to approximate the fields and potentials generated by steady state and time-varying currents. The results from this section can also be applied to the fields generated from static and oscillating magnetic dipoles. Here, we simulate a current loop using a variable number of point charges which are evenly spaced apart, where the $x$ and $y$ positions of the point charges are defined by
\begin{align}
    x\left(t\right) &= r \cos\left(\theta\left(t\right)\right), \\
    y\left(t\right) &= r \sin\left(\theta\left(t\right)\right),
\end{align}
where $r$ is the radius of the loop and $\theta(t)$ is the angle as a function of time from the $x$ axis to the point charge. The average current in a loop using point charges is given by $I=vnq/2\pi r$, where $q$ is the charge value of each point charge, $v$ is the speed of the charges, and $n$ is the number of charges used in the simulation.

For a steady state current, the angle $\theta(t)$ is
\begin{equation}
    \theta(t) = \omega t + \phi,
\end{equation}
where $\omega$ is the angular frequency of the radially accelerating point charge and $\phi$ is the phase. To ensure the point charges are equidistant from each other, the difference between point charge phases can be defined as $\Delta\phi = 2\pi/n$. The angle $\theta(t)$ for a time-varying oscillating current is given by
\begin{equation}
    \theta(t) = \frac{v_{\rm max}}{r\omega} \cos(\omega t) + \phi,
\end{equation}
where $v_{\rm max}$ is the maximum speed of the point charge and $\omega$ is the angular frequency of the current oscillations.

\begin{figure}[t]
    \centering
    \includegraphics[width=\linewidth]{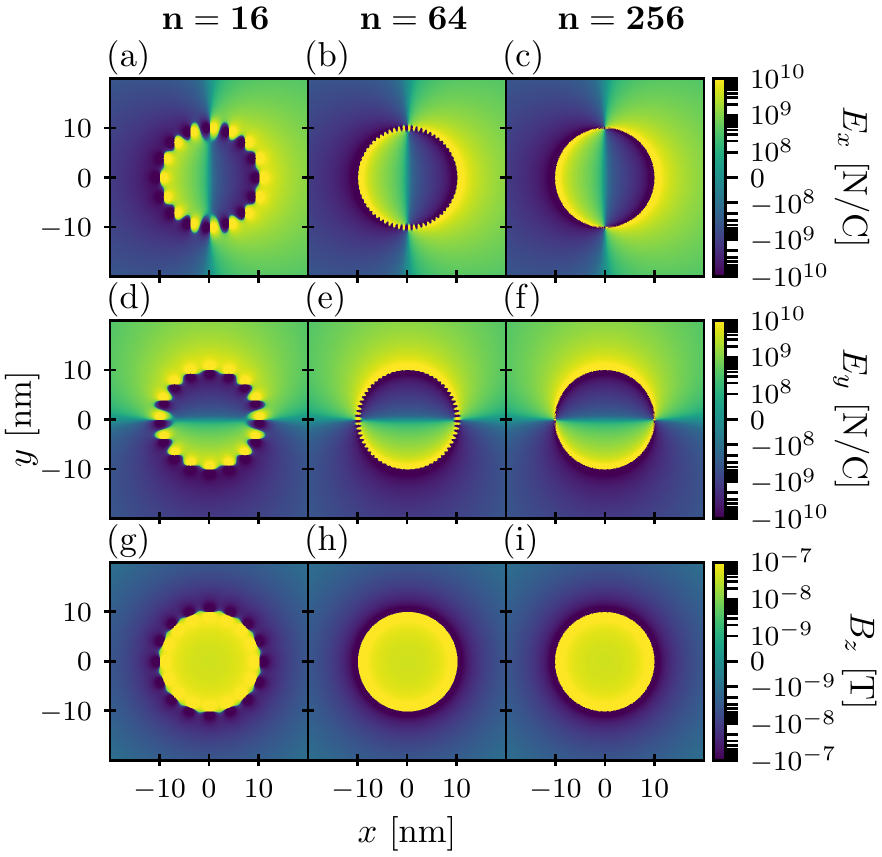}
    \caption{EM fields in the $xy$ plane generated by an approximated steady state circular current loop with a radius of 10~nm using $n$ moving point charges. From left to right, the subplots contain contain 16, 64, and 256 point charges respectively, and each simulation contains a total charge of 256e: (a)--(c) total~$E_x$; (d)--(f) total~$E_y$; (g)--(i) total~$B_z$.}
    
    \label{fig:current_xy}
\end{figure}

The calculated electric and magnetic field components of a steady state current loop with radius 10 nm are shown in Fig.~\ref{fig:current_xy} using 16, 64, and 256 point charges. The plots are along the plane of the current loop, and the simulated current is kept constant by ensuring that the sum of the charges is always 256$e$. In this simulation, the point charges are moving at a speed of 1 m/s and thus $I\simeq6.528\times10^{-10}$~A.

The accuracy of the current approximation depends on the number of point charges used in the simulation, as well as the distance between the field point being measured and the point charges. 
As shown in Fig.~\ref{fig:current_xy}, the expected circular shape of the loop becomes more defined as the number of point charges increases. The magnetic field along the center axis of the loop has a well-known analytical solution that is often derived in introductory EM courses, however the equations describing the fields off-axis require more rigorous calculus~\cite{Garrett_1951}. Our simulation allows the user to appreciate the full electric and magnetic fields generated from a current of arbitrary shape at any point in space. 

\begin{figure}
    \centering
    \includegraphics[width=\linewidth]{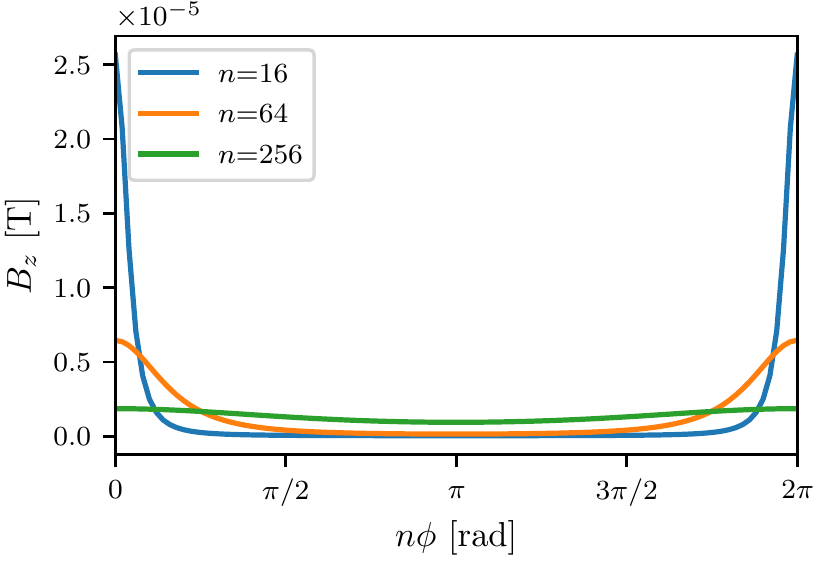}
    \caption{$B_z$ as a function of $n\phi$ at $\mathbf{r}=(9.9~\mathrm{nm}, 0~\mathrm{nm}, 0~\mathrm{nm})$, where $n$ is the number of point charges moving radially around the loop (as shown in Fig.~\ref{fig:current_xy}) and $\phi$ is the rotation of the point charges from their initial position in radians.}
    \label{fig:current_Bz}
    \vspace{10pt}
    \centering
    \includegraphics[width=\linewidth]{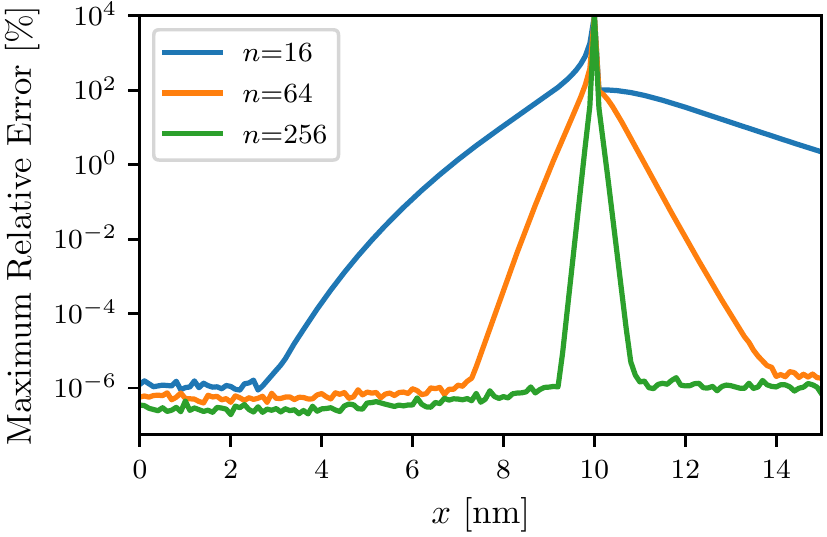}
    \caption{The maximum relative error as a function of $x$ using different amounts of point charges, where the point charges approximate a circular current loop as shown in Fig.~\ref{fig:current_xy}}
    \label{fig:current_rel_error}
\end{figure}

The amount of numerical error caused by the discretization of the current can be determined by measuring the field while rotating the point charges around the loop and observing the changes in the field. In a perfect simulation of a continuous current, the field would not change as the point charges' positions are rotated. If there are $n$ charges in the simulation, the geometry of the simulation is the same after rotating the charges by $2\pi/n$. Therefore, in Fig.~\ref{fig:current_Bz} we measure the different $B_z$ values as a function of the rotation $\phi$ at the position $\mathbf{r}=(9.9~\mathrm{nm}, 0~\mathrm{nm}, 0~\mathrm{nm})$, where $\phi\in [0, 2\pi/n]$. At this field point, the $B_z$ field varies significantly as the charges are rotated. The overall fluctuations reduce when $n$ is increased, however even with 256 point charges the field does not remain stationary. The average $B_z$ value over the full rotation for all values of $n$ converge to the same value, which is the limit using infinite point charges, representing a perfectly continuous current. 

By calculating the difference between the maximum field value over the rotation and the average field value, the maximum relative error can be determined for any point in the simulation. In Fig.~\ref{fig:current_rel_error}, we plot the maximum relative error for the simulated current loop across the $x$ axis from 0 nm to 15 nm using 16, 64, and 256 point charges. The number of point charges significantly affects the maximum relative error near the point charges at $x=10$ nm: from the origin to the point charges at $x=10$ nm, the maximum relative error becomes larger than 1\% around $x=6.862$ nm using 16 point charges, while only reaching this threshold at $x=9.752$ nm using 256 point charges. Thus, overall the point charge approximation achieves very high accuracy, attaining a maximum relative error lower than $10^{-6}$ at the origin for 256 point charges. The simulation of currents using our computational method could be used as sources in conjunction with other numerical electromagnetic simulators to investigate the interaction of EM fields with nanometer-scale structures.

\subsection{Linear Acceleration}

\begin{figure*}
    \centering
    \includegraphics[width=\linewidth]{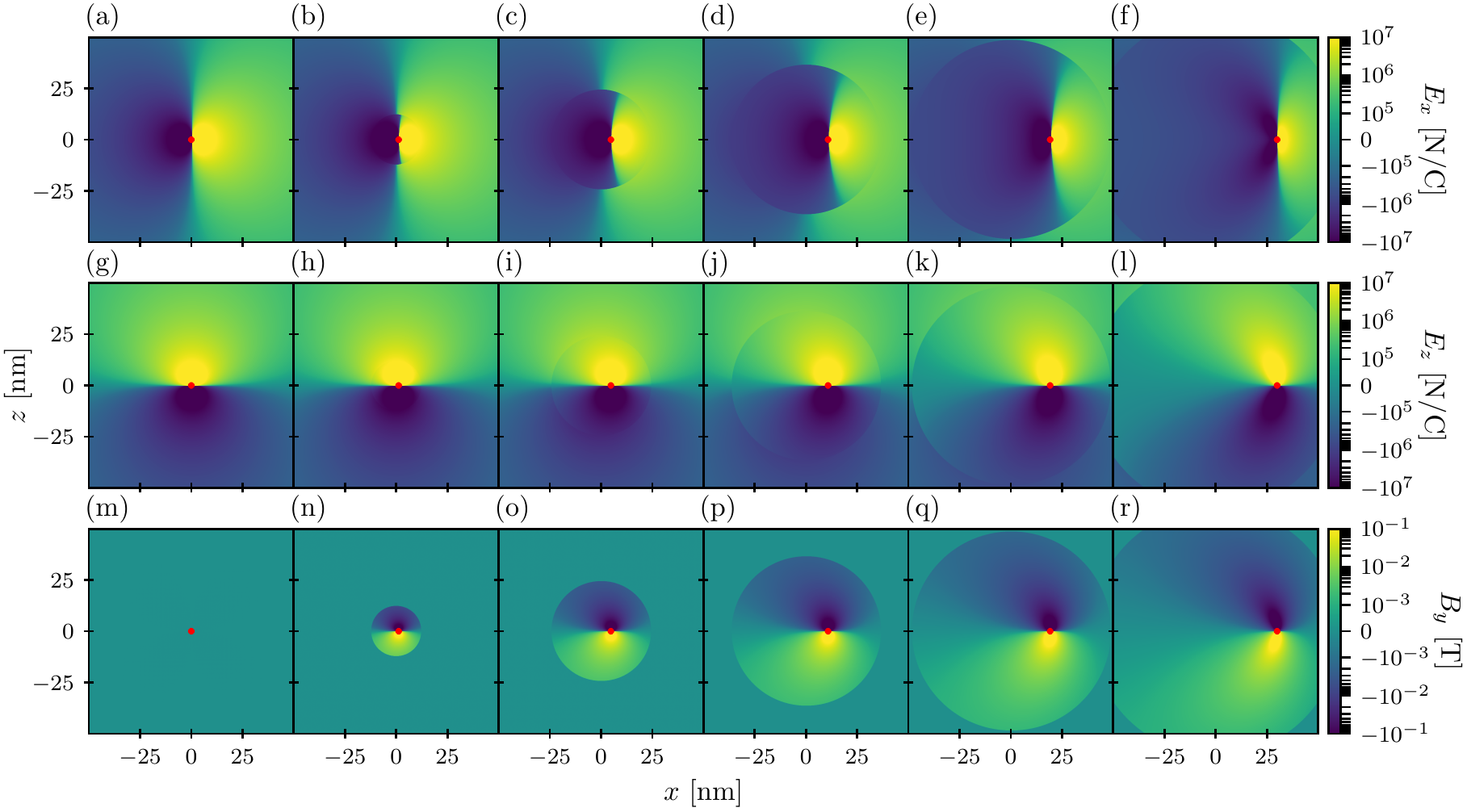}
    \caption{EM fields from a linearly accelerating positive point charge along the $x$ axis, where the red dot is the current position of the charge. The charge accelerates from $x=0$~nm while stationary and reaches a speed of 0.99$c$ at $x=30$ nm ($\tau\simeq 2.022\times 10^{-16}$~s), accelerating at approximately $1.468\times 10^{24}~\mathrm{m/s^2}$. Snapshots in the $xz$ plane starting at $t=0$~s and increasing with equal increments of $\tau/5$ in time from left to right:  (a)--(f) total $E_x$; (g)--(l) total $E_z$; (m)--(r) total $B_y$.}
    \label{fig:xz_acc_field}
\end{figure*}

The total power radiated by a point charge as it accelerates has a well-known solution given by the Larmor formula for non-relativistic point charges~\cite{F.R.S_1897}, while the Liénard's generalization of the Larmor formula is used to evaluate particles moving at relativistic speeds. Figure~\ref{fig:xz_acc_field} plots the total electric and magnetic fields of a linearly accelerating point charge at different equal increments of times. At $t=0$~s, the charge is stationary and therefore only the Coulomb field is present. As shown in the figure, when the particle begins accelerating the EM information travels outwards radially at the speed $c$. The EM waves in the radiation field are emitted in a toroidal shape about the direction of acceleration that is stretched forward, with respect to the velocity, by the factor $1/(1-\beta\cos\theta)^5$, where $\theta$ is the angle between the acceleration vector and the unit vector $\mathbf{n}$~\cite{Griffiths_2017}. The radiation generated by the linear accelerating point charge has a  similar shape to the fields produced from the oscillating single charge as it accelerates from its maximum amplitude to the trajectory's midpoint, however this acceleration is not constant with time.

\subsection{Bremsstrahlung}

\begin{figure*}
    \centering
    \includegraphics[width=\linewidth]{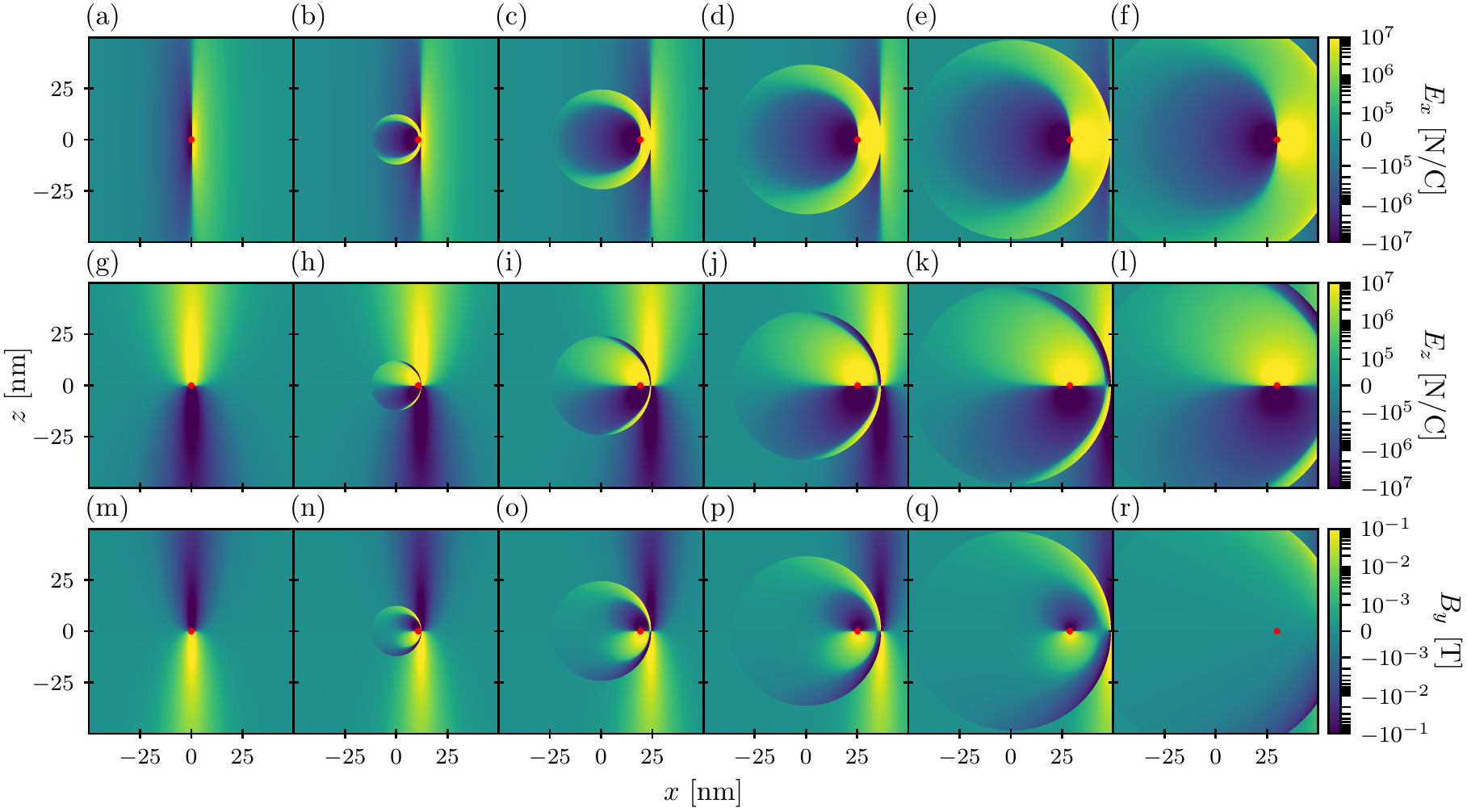}
    \caption{EM fields from a linearly decelerating positive point charge along the $x$ axis, where the red dot is the current position of the charge. The charge begins decelerating  from an initial speed of $0.99c$ at $x=0$~nm and reaches a speed of zero at $x=30$~nm ($\tau=2.022\times 10^{-16}$~s), decelerating at approximately $1.468\times 10^{24}~\mathrm{m/s^2}$. Snapshots in the $xz$ plane starting at $t=0$~s and increasing with equal increments of $\tau/5$ in time from left to right:  (a)--(f) total $E_x$; (g)--(l) total $E_z$; (m)--(r) total $B_y$.}
    \label{fig:xz_dec_field}
\end{figure*}

Bremsstrahlung, also known as ``braking radiation,'' is produced when a charged particle decelerates. Bremsstrahlung can occur when a charged particle is deflected by other charged particles as it travels through matter. Here, we plot the total electric and magnetic fields of a linearly decelerating point charge in Fig.~\ref{fig:xz_dec_field} at different equal increments in time.  At $t=0$, the charge is moving at a constant speed of 0.99$c$ so only the Coulomb field is present. At this speed, interesting relativistic effects occur in the EM fields, which are flattened in the direction perpendicular to motion. The EM fields are reduced in the forward and backwards direction by a factor $(1-\beta^2)$ relative to the field at rest, and enhanced in the perpendicular direction by a factor $1/\sqrt{1-\beta^2}$~\cite{Griffiths_2017}.  The radiation produced by the decelerating charge propagates outwards radially at the speed of light, and as the point charge's speed decreases the EM fields near the charge begin to resemble the expected static field results.

\section{Conclusions}
\label{sec:conclusions}

We have introduced a computational method to directly simulate
and visualize the EM fields and potentials generated from moving point charges, and have discussed several simulation examples where the charges follow different trajectories of motion, including sinusoidal oscillations, linear and radial acceleration, and linear deceleration. Our program, written in {\sc Python~3.8}, allows the user to specify the number of charges and their trajectories in three dimensions, and can be used to simulate steady state and time-varying currents. The simulation calculates the Coulomb, radiation, and total fields, as well as the scalar and vector potentials, at specified grid points in time by first determining the retarded time at each point. These simulations provide useful insights to gain an intuitive understanding of point charge radiation, and can be used as teaching tools for advanced undergraduate and graduate-level EM theory courses. Future work includes implementing numerical modeling methods, such as FDTD, to investigate the interactions of these generated fields with physical nanostructures. 

\section*{Acknowledgements}
This work was supported by the Natural Sciences and Engineering Research Council of Canada, Queen's University, and the Canadian Foundation for Innovation. 

\bibliography{refs}
\end{document}